%% file: main.tex
\documentclass[11pt]{article}

\usepackage{mathpazo}

\usepackage{graphicx,wrapfig}
\usepackage{lipsum}
\usepackage{amsmath,amssymb,mathrsfs}
\usepackage{hyperref}
\usepackage{url}
\usepackage{mathtools}
\usepackage{changepage}
\usepackage{multirow}
\usepackage{wrapfig}
\usepackage{subfig}
\usepackage{color}
\usepackage{epsfig,enumerate,amsmath,amsfonts,amssymb,amsthm,mathrsfs,ifpdf}
\usepackage{indentfirst,relsize}
\usepackage[numbers]{natbib}
\usepackage{setspace,graphicx}
\usepackage{latexsym}
\usepackage[all]{xy}
\usepackage[usenames,dvipsnames]{pstricks}
\usepackage{pst-grad}
\usepackage{pst-plot}
\usepackage{siunitx}

\usepackage{fullpage}

\def\Z{\ensuremath{\mathbb{Z}}}

\newcommand{\size}[1]{\left| #1 \right|}

\newcommand{\cP}{\mathcal{P}}
\newcommand{\remove}[1]{}

\newtheorem{theo}{Theorem}
\newtheorem{lem}[theo]{Lemma}

\newtheorem{cl}[theo]{Claim}

\newtheorem{defi}[theo]{Definition}
\newtheorem{remk}{Remark}
\newtheorem{obs}[theo]{Observation}
\newtheorem{prob}{Problem}

\title{Grid Obstacle Representation of Graphs}
\author{
Arijit Bishnu
\footnote{
Indian Statistical Institute, Kolkata, India
}
\and
Arijit Ghosh
\footnotemark[1]
\and
Rogers Mathew
\footnote{
Indian Institute of Technology - Hyderabad, India
}
\and
Gopinath Mishra
\footnotemark[1]
\and
Subhabrata Paul
\footnote{
Indian Institute of Technology - Patna, India
}
}

\begin{document}

\maketitle

\begin{abstract}
The \emph{grid obstacle representation}, or alternately, $\ell_1$-obstacle representation of a graph $G=(V,E)$ is an injective function $f:V \rightarrow \mathbb{Z}^2$ and a set of point obstacles $\mathcal{O}$ on the grid points of $\mathbb{Z}^2$ (where no vertex of $V$ has been mapped) such that $uv$ is an edge in $G$ if and only if there exists a Manhattan path between $f(u)$ and $f(v)$ in $\mathbb{Z}^2$ avoiding the obstacles of $\mathcal{O}$ and points in $f(V)$. This work shows that planar graphs admit such a representation while there exist some non-planar graphs that do not admit such a representation. Moreover, we show that every graph admits a grid obstacle representation in $\mathbb{Z}^3$. We also show NP-hardness result for the point set embeddability of an $\ell_1$-obstacle representation.

\end{abstract}

\input{obstacle-intro.tex}

\input{obstacle-existence.tex}

\input{obstacle-non-existence.tex}

\input{obstacle-nphard.tex}

\input{obstacle-conclusion.tex}

\bibliographystyle{alpha}
\addcontentsline{toc}{section}{Bibliography}
\bibliography{grid-obstacle}

\end{document}

%% file: obstacle-intro.tex
\section{Introduction}
\label{sec:intro}

\noindent In 2010, Alpert et al.~\cite{AlpertKL10} introduced the concept of
\emph{obstacle representation} of a graph which is closely related to
visibility graphs~\cite{Ghosh-visbook,GhoshG-acmreview-13,Goodman-handbook}. Their attempt was to represent every graph $G=(V,E)$, $\size{V}=n, \size{E}=m$,
in the Euclidean plane with a point set $P=V$ and a set $\mathcal O$ of polygonal obstacles such that for every edge $pq\in E$, $p$ and $q$ are visible in the Euclidean plane and every non-edge (a non-edge is a pair of vertices $p,q \in V$ with $pq \not\in E$) is blocked by some obstacle $o\in {\mathcal O}$ or points in $P$.
The smallest number of obstacles needed to represent a graph $G$ is called the \emph{obstacle number} of $G$ and is denoted by $obs(G)$. Clearly,
$obs(G) \leq n(n-1)/2$.

Starting with the work of Alpert et al.~\cite{AlpertKL10}, there
have been several studies \cite{Balko2015,DujmovicM15,FulekSS11,JohnsonS11,JohnsonS14,MukkamalaPP12,MukkamalaPS-largeobs-10,PachS11,Sarioz11} on existential and optimization related questions on obstacle number. In \cite{AlpertKL10}, Alpert et al.~identified some families of graphs having obstacle number $1$ and constructed graphs with obstacle number
$\Omega(\sqrt{\log n})$. Pach et al.~\cite{PachS11} showed the existence of graphs with arbitrarily large obstacle number using extremal graph theory. Dujmovi\'c et al.~\cite{DujmovicM15} proved a lower bound of $\Omega(n/ (\log \log n)^2)$ on the obstacle number. Balko et al.~\cite{Balko2015} showed that the obstacle number for general graphs is $O(n\log n)$ and for graphs with bounded chromatic number it is $O(n)$.

\subsection{Obstacle representation problem}
\label{ssec:def}

\noindent
The essence of obstacle representation of a graph is about blocking the visibility in the Euclidean plane among pairs of points whose corresponding vertices do not have an edge. The idea of ``blocking visibility''~\cite{DumitrescuPT2009,Matousek2009} between pairs of points in the Euclidean plane with a minimum set of point obstacles or blockers has resonance with the obstacle representation --- consider the obstacle representation of a graph with a finite number of vertices but no edges.
In the Euclidean plane, the shortest path and straight-line visibility are
essentially the same. To generalize the definition of obstacle representation
to other metric spaces, we replace \emph{visibility blocking} in the Euclidean
plane by \emph{shortest path blocking} in a metric space.
An obstacle representation of a graph $G= (V, E)$ in a metric space
$({\mathcal M}, \delta)$ consists of an injective mapping from $V$ to $P \subseteq {\mathcal M}$ and a set of point obstacles to be placed on points of ${\mathcal M}$.

\begin{defi}[Visibility in a metric space]\label{Def:obsvisibility}
Let $(\mathcal{M}, \delta)$ be a metric space, $\mathcal{O}\subseteq \mathcal{M}$ be the set of obstacles, 
and $S \subseteq \mathcal{M}$. Two points
$p_1, p_2 \in S$ are \emph{visible} if there exists a shortest path between $p_1$ and
$p_2$ that is not blocked by any point from $\mathcal{O} \cup S$.
\end{defi}

Observe that the shortest paths in ${\mathcal M}$ depend on the metric $\delta$, and need not be unique.

\begin{defi}[Obstacle representation problem]\label{Def:obsmetric}

	Given a graph $G = (V,E)$, a metric space $({\mathcal M}, \delta)$ and a
	point set $P \, \subseteq \mathcal{M}$, an \emph{obstacle representation} of $G$ in $({\mathcal M}, \delta)$
    consists of an injective mapping $f: V \rightarrow P$ and
    a set of obstacles ${\mathcal O}$ to be placed on points of ${\mathcal M}$,
    such that
    \begin{enumerate}
    	\item[$(i)$] for each edge $uv \in E$, $f(u)$ and $f(v)$ are mutually \emph{visible}, that is, there exists a shortest path between $f(u)$ and $f(v)$ that is not blocked by any point from $\mathcal{O} \cup f(V)$ and
	\item[$(ii)$] for each non-edge $uv \notin E$, $f(u)$ and $f(v)$ are \emph{not visible}.
	\end{enumerate}
	The minimum number of obstacles required to get an obstacle representation
    of $G$ in $(\mathcal{M}, \delta)$ is the
    \emph{$\delta$-obstacle number} of $G$ and is denoted by $\delta$-$obs(G)$.
\end{defi}

\remove{
\begin{defi}[Obstacle representation problem]\label{Def:obsmetric}
	Given a graph $G = (V,E)$, a metric space $({\mathcal M}, \delta)$ and a
	point set $P \, \subseteq \mathcal{M}$, an \emph{obstacle representation} of $G$ in $({\mathcal M}, \delta)$
    consists of an injective mapping $f: V \rightarrow P$ and
    a set of obstacles ${\mathcal O}$ to be placed on points of ${\mathcal M}$,
    such that
	$(i)$ for each edge $uv \in E$, $f(u)$ and $f(v)$ are \emph{visible} by a
	path of length $\delta(f(u),f(v))$, and
	$(ii)$ for each non-edge $uv \notin E$, $f(u)$ and $f(v)$ are \emph{not visible} by any path of length
	$\delta(f(u),f(v))$.

	The minimum number of obstacles required to get an obstacle representation
    of $G$ in $(\mathcal{M}, \delta)$ is the
    \emph{$\delta$-obstacle number} of $G$ and is denoted by $\delta$-$obs(G)$.\qed
\end{defi}
}

\begin{figure}
	\begin{center}
		\includegraphics[width = 7.0cm]{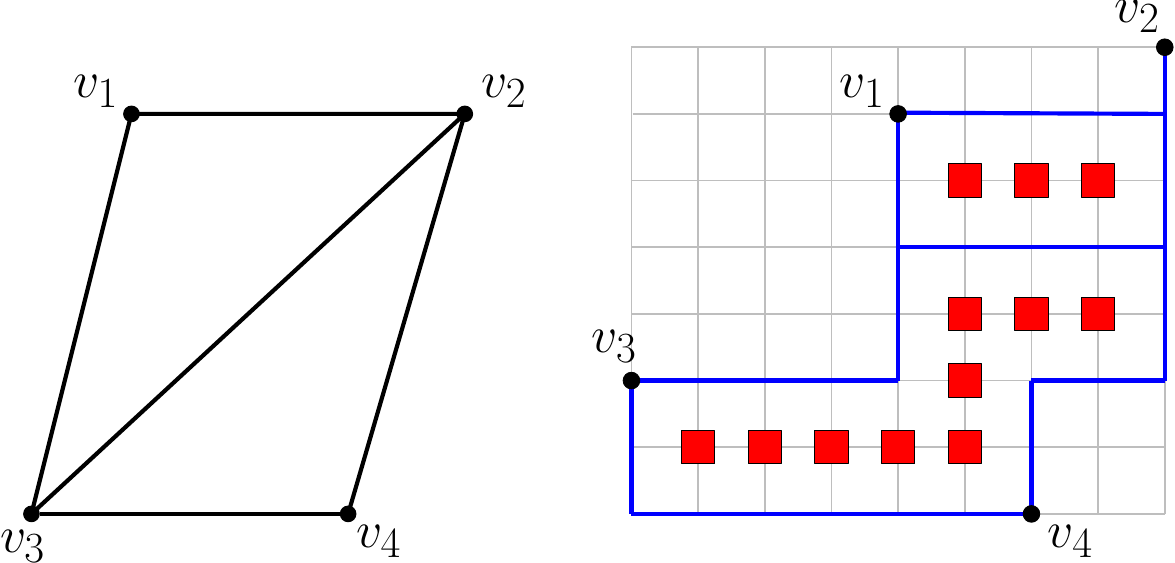}
		\caption{Grid obstacle representation of a graph; the $\ell_1$-obstacle
		representation in $\mathbb{Z}^{2}$ is on the right for the graph on the left side.}
		\label{fig:graph}
	\end{center}
\end{figure}	

There is a minor technical point here though. In a discrete metric space like
$\mathbb{Z}^{d}$, a path is defined as a path in the unit distance graph.
Note that points corresponding to the vertices of the graph can also act as an
obstacle. In the above definition, the obstacle representation is
influenced by the metric space $(\mathcal{M}, \delta)$ and the
obstacles. In the light of the above definition, Alpert et al.'s~\cite{AlpertKL10} representation is an $(\mathbb{R}^2, \ell_2)$
representation with polygonal obstacles. In this paper, we restrict ourselves to
$(\mathbb{Z}^{d}, \ell_1)$ with point obstacles. We term this representation as the \emph{grid obstacle representation}, or alternately,
\emph{$\ell_1$-obstacle representation} of $G$.
The \emph{grid obstacle number} of $G$ is the minimum number of obstacles needed for a grid obstacle representation of $G$. For the
$\ell_1$-metric, the shortest path between two points is not
unique. So, in the obstacle representation, we need to block all such
shortest paths. Figure \ref{fig:graph} illustrates the $\ell_1$-obstacle
representation on $\Z^2$. \remove{For more examples on different types of graphs, refer to Appendix \ref{app:examplesgridrepresentation}. We have observed interesting behavior about the existence of grid obstacle representation. The presence of a certain number of $C_4$, a cycle of length $4$, is acting like a forbidden structure for existence of grid obstacle representation for graphs with \emph{high} number of edges. One can show that there exists a $C_4$-free graph on more than $20$ vertices having at least $8n-19$ edges that does not admit a grid obstacle representation. At the same time, graphs having $C_4$ for all four tuple of vertices, admits grid obstacle representation.}

Starting with this definition of ours~\cite{bishnu-arXiv-grid}, there has already been substantial interest in this particular representation, as evidenced from the works in~\cite{Biedl2018,GeodesicBose,JanosPach2016}.

\subsection{Our contribution}
\label{ssec:contrib}
\noindent
Apart from introducing a new obstacle representation of a graph, we deduce several existential and algorithmic results on this new representation.
Section~\ref{sec:embedding} shows that planar graphs admit $\ell_1$-obstacle representation in grids of size $O(n^4) \times O(n^4)$, whereas, in Section~\ref{app:embedding3D}, we show that every graph admits a grid obstacle representation in $\mathbb{Z}^3$.
In Section~\ref{sec:nonexistence}, we show the existence of graphs that do not admit grid obstacle representation in $\mathbb{Z}^2$. On the algorithmic side, we show a hardness result for the point set embeddability of an $\ell_1$-obstacle representation in Section~\ref{app:hardness}. Our work poses several interesting existential and algorithmic questions regarding $\ell_1$-obstacle representability, and more generally, geodesic representation of graphs in different metric spaces.

%% file: obstacle-existence.tex
\section{Existential results}

\subsection{Planar graphs in $\mathbb{Z}^2$}
\label{sec:embedding}
\noindent In this section, we show that every planar graph admits a representation in $(\mathbb{Z}^2, \ell_1)$. To this end, we use results on straight-line embeddings of planar graphs on grids. By straight-line embedding of a planar graph $G=(V,E)$ on a grid, we mean a planar embedding of $G$ where the vertices lie on grid points and edges are represented by straight lines joining the vertices. We use the straight-line embedding of a triangulated planar graph with $n$ vertices on an $O(n) \times O(n)$ grid due to~\cite{Farysseix1990,Schnyder1990}.


\begin{theo}[\cite{Farysseix1990,Schnyder1990}] \label{theo:Schnyder}
Each planar graph with $n\geq 3$ vertices has a straight-line embedding on an $(n-2) \times (n-2)$ grid.
\end{theo}

In this embedding, let $c$ be the minimum distance between the set of all grid points and the set of all embedded edges.
Let $A$ be the corresponding grid point and $\overline{BC}$ be the corresponding edge that contributes to the minimum distance. Consider $\triangle ABC$. Its area $\Delta$ is $\frac{c \cdot a}{2}$, where $a$ is the length of $\overline{BC}$. Now, since the embedding is on an $(n-2)\times (n-2)$ grid, $a < \sqrt{2}n$. Also the area $\Delta$ is $\geq \frac{1}{2}$ because the absolute value of the determinant\footnote{$\begin{vmatrix}
1 & x_{A} & y_{A} \\ 
1 & x_{B} & y_{B} \\
1 & x_{C} & y_{C}
\end{vmatrix} $} corresponding to the doubled area of the triangle is at least $1$ as $A$, $B$ and $C$ have integral coordinates and $A \not\in \overline{BC}$. So, $2 \Delta$ is a positive integer and $2\Delta \geq 1$. Hence, we have $c = 2 \Delta/a >  2 \cdot 0.5 / \sqrt{2}n = 1 / \sqrt{2}n$. Now, if we ``blow up'' or refine the grid uniformly ${\mathcal C} \cdot n$ times, where $\mathcal{C} \gg \sqrt{2}$ is a constant, then $c$ will be at least $\mathcal{C}/\sqrt{2}$. Hence, we have the following lemma.

\begin{lem}\label{lem:stlineOn2}
Every planar graph admits a straight-line embedding on an $O(n^2)\times O(n^2)$ unit grid such that the distance between a vertex $v$ and a straight-line edge $e$ not containing $v$ is greater than some constant ${\mathcal C}$. In this $O(n^2) \times O(n^2)$ grid, any two vertices of $G$ are at least $\Omega(n)$ apart, and an edge and a vertex are at least at a distance $\mathcal{C}$ apart.
\end{lem}

Using this result, we can show the existence of an obstacle representation of planar graph in $(\mathbb{Z}^2,\ell_1)$. The idea is the following:
\begin{description}
\item[1.] Obtain a straight-line embedding of a planar graph
as in Lemma~\ref{lem:stlineOn2}.
\remove{
in $O(n^2)\times O(n^2)$ grid such that a vertex $v$ is far away from a straight-line edge $e$ not containing $v$.}

\item[2.] Each vertex $v$ has around it an $\epsilon$-box $B^{\epsilon}(v)$, a square box with sides of length $\epsilon$,  such that
\begin{itemize}
  \item for two distinct vertices $u$ and $v$, $B^{\epsilon}(v) \cap B^{\epsilon}(u) = \emptyset$,
  \item the minimum distance between two $\epsilon$-boxes $B^{\epsilon}(u)$ and $B^{\epsilon}(v)$ is large (say $q_1$), and
  \item the minimum distance between an $\epsilon$-box $B^{\epsilon}(v)$ and a straight-line edge $e$ not containing $v$ is also adequate (at least $q_2$).
\end{itemize}

\item[3.] Consider a $\delta$-tube $T^{\delta}(e)$ (where $T^{\delta}(e)$ is the Minkowski sum of the embedded edge $e$ and a disk of radius $\delta$) around each straight-line edge $e$ such that
\begin{itemize}
  \item for each pair of straight-line edges $e_1$ and $e_2$ that do not share a common vertex, we have $T^{\delta}(e_1) \cap T^{\delta}(e_2)=\emptyset$,
  
  \item for each pair of distinct straight-line edges $e_1$ and $e_2$ sharing a common vertex $v$, we have $(T^{\delta}(e_1) ~ \cap ~ T^{\delta}(e_2)) \subset B^{\epsilon}(v)$, and

\item for a straight-line edge $e$ and a vertex $v \not\in e$, we have $B^{\epsilon}(v)\cap T^{\delta}(e)=\emptyset$.

\end{itemize}

\item[4.] Refine the grid in such a way that we can convert the straight-line edge $e$ into a Manhattan path that lies inside $T^{\delta}(e)$.
\end{description}



\begin{figure}
	\begin{center}
		\includegraphics[width=7cm]{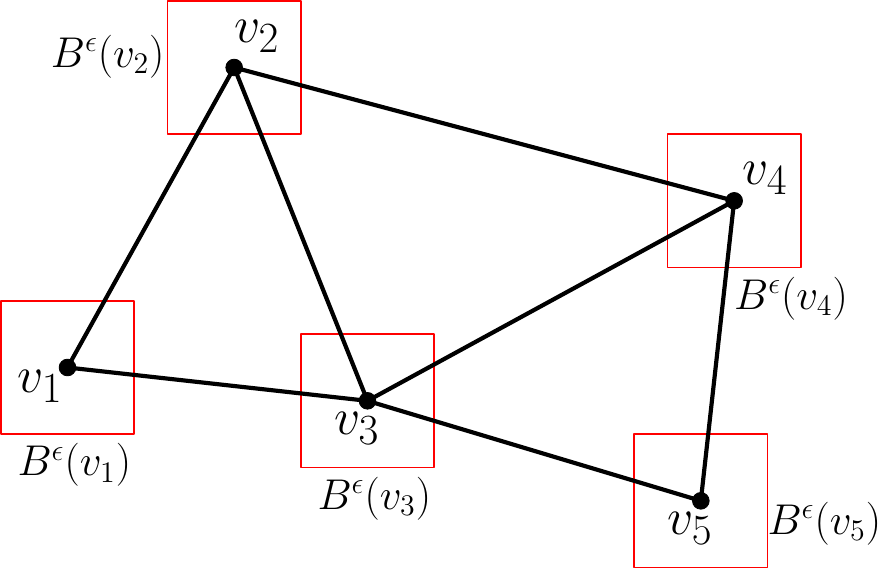}
		\caption{The $\epsilon$-box for each vertex.}
		\label{fig:epsilonbox}
	\end{center}
\end{figure}

\begin{figure}[!htb]
\centering

\begin{minipage}{0.5\textwidth}
\centering
\includegraphics[height=.40\linewidth]{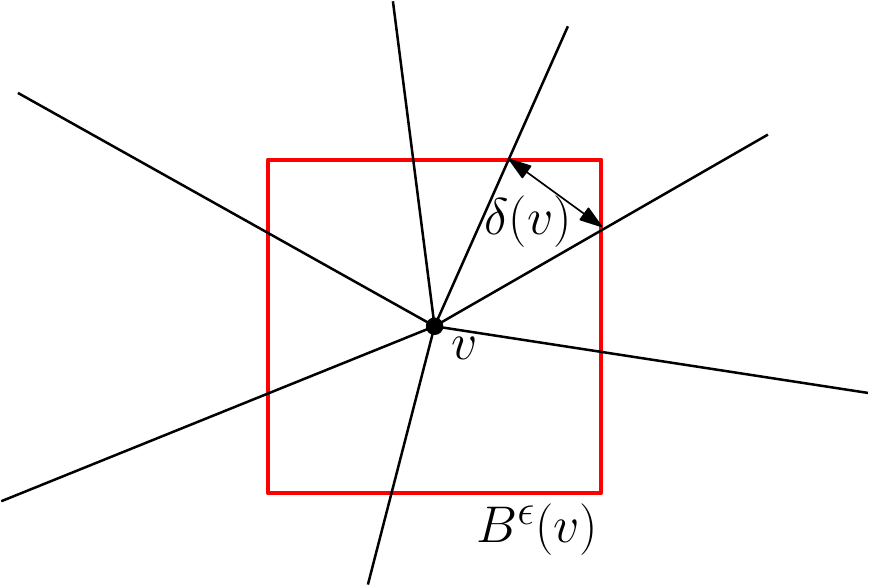}
\caption{Illustration of $\delta(v)$}
\label{fig:deltav}
\end{minipage}%
\begin{minipage}{0.5\textwidth}
\centering
\includegraphics[height=.40\linewidth]{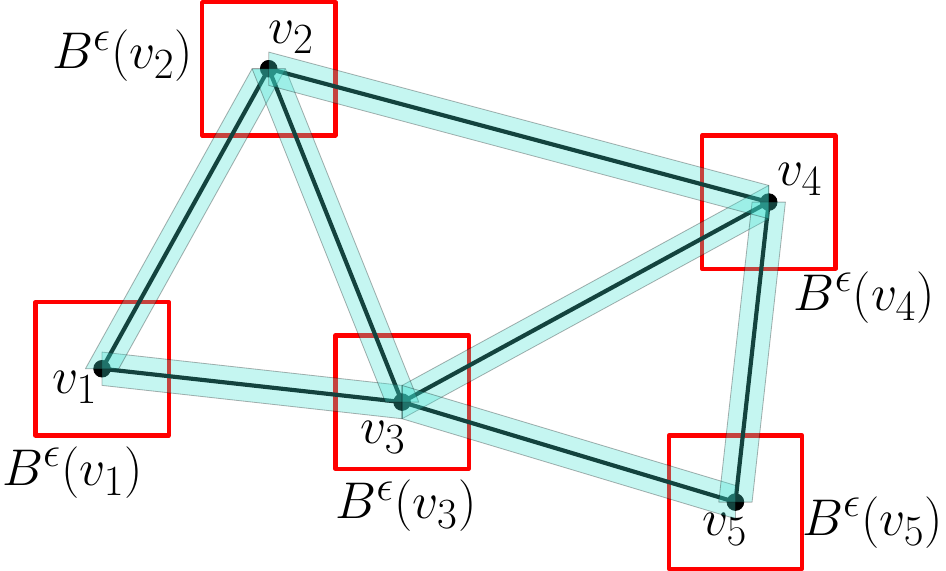}
\caption{Illustration of $T^{\delta}(e)$. The grid will be suitably refined so that some Manhattan path corresponding to every edge $e$ lies inside $T^{\delta}(e)$.}
\label{fig:tubularregion}
\end{minipage}
\end{figure}

Now, in this embedding, consider $B^{\epsilon}(v)$ as shown in Figure~\ref{fig:epsilonbox}, with $\epsilon \ll \mathcal{C}$, where $\mathcal{C}$ is the constant defined in Lemma \ref{lem:stlineOn2}. The length $\epsilon$ is chosen in such a way that $q_1 = \Omega(n - \epsilon) = \Omega(n)$ and $q_2 = (\mathcal{C}-\sqrt{2}\epsilon)$ is an adequately large constant (to be fixed as per Observation~\ref{obs:tubeproperty}). Let us consider a vertex $v$ with $\mbox{deg}(v)>1$. The straight-line edges that contain $v$ cut $B^{\epsilon}(v)$. Let $\delta(v)$ be the minimum Euclidean distance between consecutive intersection points of $B^{\epsilon}(v)$ and straight-line edges containing $v$ as shown in Figure \ref{fig:deltav}. Let $\delta < \frac{1}{10} \min\limits_{v \in V}\{\delta(v)\}$. Consider the tubular region around a straight-line edge $e$ of length $\delta$ as shown in Figure \ref{fig:tubularregion}. Let the region be denoted by $T^{\delta}(e)$. The choice of $\delta$ guarantees the following observations about $B^{\epsilon}(v)$ and $T^{\delta}(e)$.

\remove{
The length $\epsilon$ is chosen in such a way that distance between two $\epsilon$-boxes $B^{\epsilon}(u)$ and $B^{\epsilon}(v)$ is large, that is $(\mathcal{C}-\sqrt{2}\epsilon)$ is large, and distance between an $\epsilon$-box $B^{\epsilon}(v)$ and a straight-line edge $e$ not containing $v$ is also large, that is $(\mathcal{C}-\frac{\epsilon}{\sqrt{2}})$ is large.
Let $\mbox{deg}(v)$ denote the degree of a vertex $v$  and $D=\max\limits_{v\in V}\{\mbox{deg}(v)\}$.}

\remove{The distance is measured along the boundary of $B^{\epsilon}(v)$ as shown in Figure \ref{fig:deltav}. The width $\delta$ of the $\delta$-tubes is chosen so that it satisfies the inequality $D (\delta(v) + 2 \delta) \leq 4 \epsilon$, where $4\epsilon$ is the circumference of $B^{\epsilon}(v)$.}

%
%


\begin{obs}\label{obs:tubeproperty}
For the particular choices of $\epsilon$ and $\delta$, the following properties of $T^{\delta}(e)$ and $B^{\epsilon}(v)$ are true: 
\begin{itemize}
\item[$(i)$] for each pair of straight-line edges $e_1$ and $e_2$ that do not share a common vertex, $T^{\delta}(e_1) \cap T^{\delta}(e_2)=\emptyset$,
\item[$(ii)$] for each pair of distinct straight-line edges $e_1$ and $e_2$ sharing a common vertex $v$, $(T^{\delta}(e_1) \cap T^{\delta}(e_2)) \subset B^{\epsilon}(v)$, i.e., there is no intersection between $\delta$-tubes outside the $\epsilon$-boxes; 
\item[$(iii)$] for a vertex $v$ and a straight-line edge $e$ not containing $v$, $B^{\epsilon}(v)\cap T^{\delta}(e)=\emptyset$.

\end{itemize}
\end{obs}

%
%
%

Once we have fixed $\epsilon$ and $\delta$, we then refine the grid further so that the width of the tube around each edge becomes sufficiently large. To do this refinement, we need to bound the value of $\delta$ which is done in the following lemma.

\begin{lem}\label{lem:deltavalue}
The value of $\delta$ is equal to $\mathcal{C'} /n^2 $, for some constant $\mathcal{C'}$.
\end{lem}
\begin{proof}
Let minimum $\delta (v)$ be achieved by two straight line edges $vu$ and $vw$. Let $A$ and $B$ be the two points where the boundary of $B^{\epsilon}(v)$ intersects with $vu$ and $vw$, respectively. Hence, $\delta(v)={AB}$. Now, if the $\angle AvB$ is more than $90\si{\degree}$, then $|AB| > |vB| \geq \epsilon/2.$ Otherwise, the following two cases arise.

\begin{figure}[!h]
\begin{center}
\includegraphics[width=5cm]{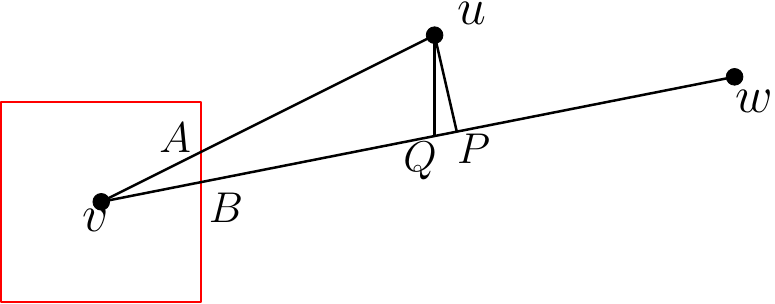}
\caption{Fixing the value of $\delta$ on the $O(n^2) \times O(n^2)$ grid\\.}
\label{fig:deltavalue}
\end{center}
\end{figure}


\noindent\textbf{Case 1:}
Assume that $A$ and $B$ are on the same side of $B^{\epsilon}(v)$ as shown in Figure \ref{fig:deltavalue}. Let $uP$ be the perpendicular on $vw$ and $uQ$ be parallel to $AB$. Note that $\size{vA} \geq \epsilon/2$. By Lemma \ref{lem:stlineOn2}, $\size{uP} > \mathcal{C}$. So, clearly $\size{uQ} > \mathcal{C}$. Since the grid is of size $O(n^2)\times O(n^2)$, we have $\size{vu} < \mathcal{C}_1 n^2$, for some constant $\mathcal{C}_1$. Now, as $\triangle vAB$ and $\triangle vuQ$ are similar, we have,
$$\frac{\size{AB}}{\size{vA}}= \frac{\size{uQ}}{\size{vu}} \Rightarrow \size{AB} > \frac{ \epsilon \mathcal{C}}{2\mathcal{C}_1}\cdot \frac{1}{n^2}$$

\begin{figure}[!h]
\begin{center}
\includegraphics[width=4cm]{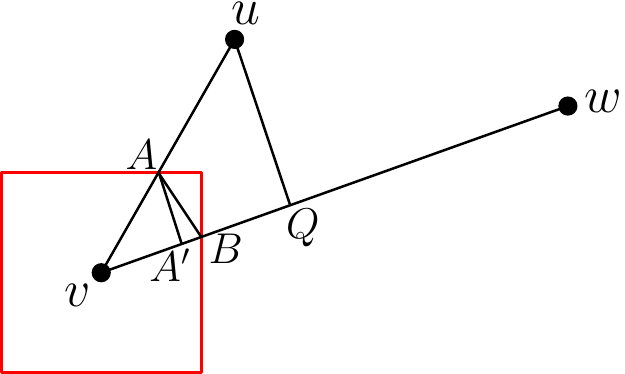}
\caption{Fixing the value of $\delta$ on the $O(n^2) \times O(n^2)$ grid.}
\label{fig:deltavalue3}
\end{center}
\end{figure}
 
\noindent\textbf{Case 2:} Assume that $A$ and $B$ are on two consecutive sides of $B^{\epsilon}(v)$, as shown in Figure \ref{fig:deltavalue3}. Let $AA'$ and $uQ$ be two perpendiculars on $vw$. In this case, $\size{AB} > \size{AA'}$. Using the similarity of $\triangle vAA'$ and $\triangle vuQ$, we have $\size{AA'} > \mathcal{C}_2 /n^2$, for some constant $\mathcal{C}_2$. Since $\size{AB} > \size{AA'}$, we have $\delta(v)> \mathcal{C}_2 /n^2$.

Now by choosing an appropriate constant $\mathcal{C'}$, we have $\delta = \mathcal{C'}/n^2 $ for all cases.
\end{proof}



Now refine the grid sufficiently until the length of each grid edge becomes $\delta /100$, i.e., $\mathcal{C'}/ (100 n^2)$. Thus we have a straight-line embedding of a planar graph on a refined grid of size $O(n^4)\times O(n^4)$. Moreover, in this embedding, there are enough grid points within $T^{\delta}(e)$ to convert the edge $e$ into a Manhattan path inside $T^{\delta}(e)$. Next, we convert the straight-line edge $e$ connecting $u$ and $v$ into a Manhattan path $M(e)$ between $u$ and $v$ in the refined grid. The existence of such a Manhattan path is guaranteed by the following lemma.
\begin{lem}\label{lem:stline_to_Manhat}
Let $e$ be a straight-line edge connecting $u$ and $v$ and the length of the refined grid edge be $\mathcal{C'} /100 n^2$. Then there exists a Manhattan path $M(e)$ connecting $u$ and $v$ such that $M(e)$ lies inside $T^{\delta}(e)$.
\end{lem}
\begin{proof}
Without loss of generality, consider the edge to be of positive slope. In the refined grid, consider the grid cells that are being intersected by the straight-line edge $e$. Since $e$ is a straight-line, two consecutive grid cells that are being intersected by $e$ have the following property: the second grid cell is either to the north, to the east or to the north-east corner of the previous grid cell as shown in Figure~\ref{fig:edge_to_path}.
\begin{figure}[!h]
\begin{center}
\includegraphics[width=8cm]{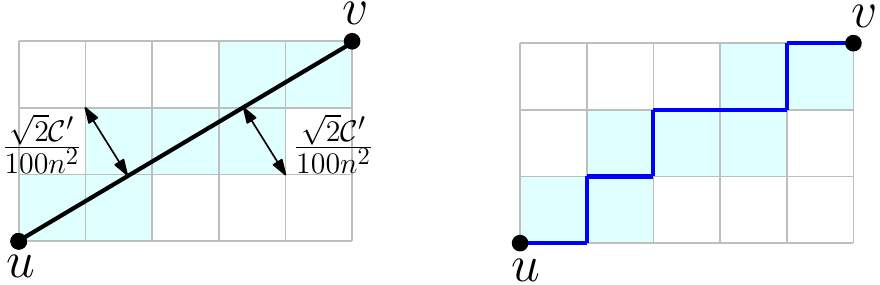}
\caption{Conversion of a straight-line edge into a Manhattan path}
\label{fig:edge_to_path}
\end{center}
\end{figure}
Consider a Manhattan path $M(e)$ from $u$ to $v$ that lies inside the union of those (closed) grid cells. Any point on $M(e)$ is within a distance of $\sqrt{2} \mathcal{C'} /100 n^2$ from the straight-line edge $e$ because each point on a grid cell that is being intersected by the straight-line edge $e$ is within the same distance from $e$. Hence, $M(e)$ lies inside $T^{\delta}(e)$.
\end{proof}

By the above lemma, we have obtained an embedding of the planar graph on a refined grid of size $O(n^4) \times O(n^4)$, where each edge $e$ is represented by a Manhattan path $M(e)$. Also the conversion of a straight-line edge $e$ into a Manhattan path $M(e)$ is done in such a way that it avoids the corner points of $B^{\epsilon}(v)$. Suppose that a Manhattan path $P$ passes through a corner point, say $x$, of $B^{\epsilon}(v)$. Consider the grid points $p$ and $q$ on $P$, such that $p$ lies before $x$, $q$ lies after $x$, they differ in both coordinates and the distance between $p$ and $q$ is the smallest possible. Modify the Manhattan path $P$ by replacing the part between $p$ and $q$ by some other Manhattan path between $p$ and $q$ that does not go through $x$.

\begin{figure}[!h]
\begin{center}
\includegraphics[width=8cm]{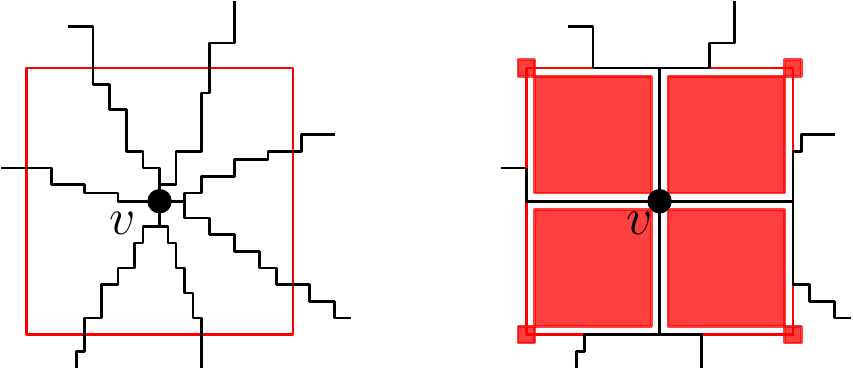}
\caption{Modification of $M(e)$ inside $B^{\epsilon}(v)$}
\label{fig:modification}
\end{center}
\end{figure}

Note that two Manhattan paths inside $B^{\epsilon}(v)$ may have non-empty intersection which might lead to a Manhattan path between two vertices of a non-edge. So, in order to block all the non-edges, we modify the Manhattan paths inside $B^{\epsilon}(v)$ and add obstacles inside $B^{\epsilon}(v)$ in the way as shown in Figure~\ref{fig:modification}. While entering $B^{\epsilon}(v)$, if an $M(e)$ intersects the horizontal (vertical) boundary of $B^{\epsilon}(v)$, the path is altered to travel along the same horizontal (vertical) boundary of $B^{\epsilon}(v)$ to intersect the vertical (horizontal) grid line through $v$, and then follow the vertical (horizontal) grid line to $v$. The new path that consists of
$M(e)$ outside $B^{\epsilon}(v)$ and the altered path inside $B^{\epsilon}(v)$ is also a Manhattan path.
 We do this modification for all $B^{\epsilon}(v)$s. Finally, we place obstacles on the four corner points of $B^{\epsilon}(v)$ and on all the grid points inside $B^{\epsilon}(v)$ except the grid points on the boundary of $B^{\epsilon}(v)$ and the vertical and horizontal line containing $v$. This is shown in Figure \ref{fig:modification}. We also place obstacles on each empty grid point outside $B^{\epsilon}(v)$s. We show that this embedding is an obstacle representation of the planar graph on an $O(n^4)\times O(n^4)$ grid.


Let us color all the paths present in the embedding into two colors, \emph{green} ({\sf G}) and \emph{blue} ({\sf B}). The portion of a path that is inside $B^{\epsilon}(v)$, for some $v\in V$, is colored green and all the remaining portion of the path is colored blue. We apply this coloring technique for all the paths in the embedding. For this coloring, we have the following lemma.
\begin{lem}\label{lem:GBG}
Each Manhattan path that starts and ends at some vertices is of the form {\sf GBG}, i.e., the path starts with a green portion, has a following blue portion and ends with another green portion.
\end{lem}

\noindent{\bf Proof.} Let $\cP$ be a path in the embedding that starts at $u$ and ends at $v$.
\begin{figure}[h]
	\begin{center}
		\includegraphics[width=4.0cm]{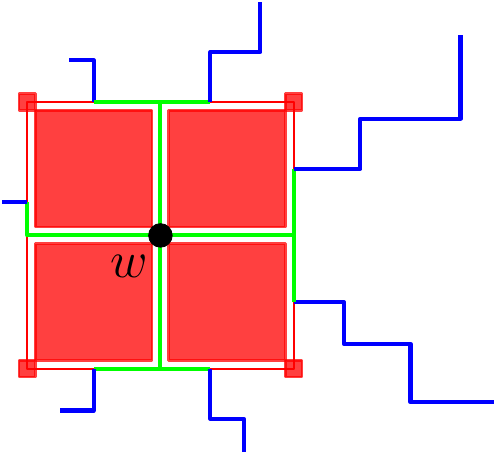}
		\caption{Path of the form GBGBG is not Manhattan}
		\label{fig:GBG}
	\end{center}
\end{figure}
Since the starting and ending portion of $\cP$ belongs to $B^{\epsilon}(u)$ and $B^{\epsilon}(v)$ respectively, both the end portions will be green. If there is only one blue portion in $\cP$ between these starting and ending green portions, then $\cP$ is of the form {\sf GBG}. Hence, without loss of generality, assume that there are two blue portions. All the blue portions have to be disjoint because of Observation~\ref{obs:tubeproperty} and Lemma~\ref{lem:stline_to_Manhat}. So, there must be a green portion between these two blue portions. Hence, the path is of the form {\sf GBGBG}.


Let the middle green portion of $\cP$ belong to $B^{\epsilon}(w)$. Notice that according to our definition, $w$ also acts as an obstacle. By the placement of the obstacles, including the corners of $B^{\epsilon}(w)$, it is clear that both the blue portions must touch the same side of $B^{\epsilon}(w)$. See Figure~\ref{fig:GBG}. The path can not be a straight Manhattan path also because of the corner obstacles. Hence, $\cP$ can not be a Manhattan path and each Manhattan path is of the form {\sf GBG}.
\qed

Using this lemma, we show that each edge $uv\in E$ corresponds to a Manhattan path between $u$ and $v$.

\begin{lem}
There is an edge $uv\in E$ if and only if there is a Manhattan path between $u$ and $v$ in the embedding.
\end{lem}
\begin{proof}
By Lemma \ref{lem:stline_to_Manhat}, it is clear that each edge $uv$ in the planar graph is represented by a Manhattan path $M(uv)$ connecting $u$ and $v$.

Conversely, in the embedding, let us assume that there is a Manhattan path $\cP$ connecting $u$ and $v$. By Lemma \ref{lem:GBG}, $P$ is of the form GBG. Note that all the blue portions lie outside the $\epsilon$-boxes and are disjoint because of Observation~\ref{obs:tubeproperty}. Hence, the blue portion of $\cP$ is exactly a blue portion of some $M(e)$. Let us now assume that the two ends of this blue portion of $M(e)$ touch two boxes $B^{\epsilon}(u)$ and $B^{\epsilon}(v)$, respectively. This implies that $e$ is incident to both $u$ and $v$. Because of Observation~\ref{obs:tubeproperty}, if a blue portion of $M(e)$ touches some $B^{\epsilon}(v)$, then $e$ is incident to $v$. Hence, the Manhattan path $\cP$ connecting $u$ and $v$ in the embedding represents the edge $uv\in E$.
\end{proof}

Hence, we have the following theorem.

\begin{theo}\label{theo:planar}
Every planar graph admits a $(\mathbb{Z}^{2}, \ell_1)$ obstacle representation on a grid of size $O(n^4)\times O(n^4)$.
\end{theo}

\remove{
\begin{remk}
In proving Theorem~\ref{theo:planar}, we came up with an embedding where each edge $e$ is represented by a Manhattan path $M(e)$ and each vertex $v$ is contained inside a square $B^{\epsilon}(v)$. A close variant of such a representation is studied in~\cite{Fossmeier1997}. To use this setting for grid obstacle representation of a planar graph, we have to place a vertex on a grid point inside the corresponding rectangle. But this need not produce a grid obstacle representation as there is no guarantee that all paths are Manhattan.
\end{remk}

\begin{remk}
In the proof of Theorem \ref{theo:planar}, as an intermediate step, we have come up with an embedding of a planar graph where each edge $e$ is represented by a Manhattan path $M(e)$ and each vertex $v$ is contained inside of a square, namely $B^{\epsilon}(v)$. A relevant representation is studied in \cite{Fossmeier1997} where the vertices are represented by rectangles and the edges are composed of a vertical and of a horizontal segment joining a vertical side of a rectangle with a horizontal side of another rectangle. Using this setting, to achieve a grid obstacle representation of a planar graph, we have to place a vertex on a grid point inside the corresponding rectangle. But this need not produce a grid obstacle representation of the planar graph as some paths fail to be Manhattan.
\end{remk}
}

\subsection{Embedding in $\mathbb{Z}^3$}
\label{app:embedding3D}
\noindent
In this section, we show the existence of an $\ell_1$-obstacle representation for any graph in $\Z^3$. The proof is also constructive and is based on the following theorem by Pach \emph{et al.}~\cite{PachTT97}.

\begin{theo}[\cite{PachTT97}] \label{theo:Pach3dim}
For every fixed $r\geq 2$, any $r$-colorable graph with $n$ vertices has a straight-line embedding in $\mathbb{Z}^3$ on a grid of size $O(r) \times O(n) \times O(rn)$ such that no two edges intersect.
\end{theo}

First, we construct a straight-line embedding of an $r$-colorable graph in a refined grid such that the distance between a vertex $v$ and a straight-line edge $e$ not containing $v$ is \emph{sufficiently large}. To do that, we give a bound on the minimum distance of a grid point from a straight-line edge in the embedding that is obtained from Theorem \ref{theo:Pach3dim}. Clearly, this distance is smaller than or equal to the distance of a vertex from a straight-line edge. Using similar arguments as in section \ref{sec:embedding}, we get that the minimum distance of a grid point from a straight-line edge is $O(1/rn)$, where $r$ is the chromatic number of the graph. Now we blow up the grid uniformly by a factor of $\mathcal{C} \cdot rn$ such that the distance between a vertex $v$ and a straight-line edge $e$ not containing $v$ becomes greater than the constant $\mathcal{C}$. Hence, we have the following lemma.

\begin{lem}\label{lem:stlineon3dim}
Every $r$-colorable graph admits a straight-line embedding on a grid of size $O(r^2n)\times O(rn^2)\times O(r^2n^2)$ such that two vertices are apart by a distance of at least $\mathcal{C}$ and a vertex and an edge are also apart by a distance of at least $\mathcal{C}$, where $\mathcal{C}$ is a constant.
\end{lem}

The idea of the proof is similar to the proof presented in section \ref{sec:embedding}. The idea is as follows:

\begin{description}
\item[1.] Obtain a straight-line embedding of an $r$-colorable graph as in Lemma~\ref{lem:stlineon3dim}.

\item[2.] Consider an $\epsilon$-cube $C^{\epsilon}(v)$, a cube of length $\epsilon$, around each vertex $v$ such that
\begin{itemize}
  \item for two distinct vertices $u$ and $v$, $C^{\epsilon}(v) \cap C^{\epsilon}(u) = \emptyset$,
  \item distance between two $\epsilon$-cubes $C^{\epsilon}(u)$ and $C^{\epsilon}(v)$ is large (say $q_1$), and
  \item distance between an $\epsilon$-cube $C^{\epsilon}(v)$ and a straight-line edge $e$ not containing $v$ is also adequate (at least $q_2$).
\end{itemize}

\item[3.] Consider a $\delta$-tube $T^{\delta}(e)$ (where $T^{\delta}(e)$ is the Minkowski sum of the embedded edge $e$ and a ball of radius $\delta$) around each straight-line edge $e$ such that
\begin{itemize}
  \item for each pair of straight-line edges $e_1$ and $e_2$ that do not share a common vertex, $T^{\delta}(e_1) \cap T^{\delta}(e_2)=\emptyset$,
  
  \item for each pair of distinct straight-line edges $e_1$ and $e_2$ sharing a common vertex $v$, we have $(T^{\delta}(e_1) ~ \cap ~ T^{\delta}(e_2)) \subset C^{\epsilon}(v)$, and
  \item for a vertex $v$ and a straight-line edge $e$ not containing $v$, $C^{\epsilon}(v)\cap T^{\delta}(e)=\emptyset$.
\end{itemize}

\item[4.] Refine the grid in such a way that we can convert the straight-line edge $e$ into a Manhattan path that lies inside $T^{\delta}(e)$.
\end{description}

Now, let us consider $\epsilon \ll \mathcal{C}$, where $\mathcal{C}$ is the constant given in Lemma \ref{lem:stlineon3dim}. The value of $\epsilon$ is chosen in such a way that both $q_1=(\mathcal{C} - \epsilon)$ and $q_2=(\mathcal{C}- \sqrt{2}\epsilon)$ are adequately large constants. Next, we fix $\delta$ to be a constant such that $\delta < \frac{1}{10} \min\limits_{v\in V}\{\delta(v)\}$, where $\delta(v)$ is the minimum Euclidean distance between consecutive intersection points of $C^{\epsilon}(v)$ and straight-line edges containing $v$. The choice of $\delta$ guarantees the following observations about $C^{\epsilon}(v)$ and $T^{\delta}(e)$.

\begin{obs}\label{obs:tubeproperty_on_3}
For the particular choices of $\epsilon$ and $\delta$, the following properties of $T^{\delta}(e)$ and $C^{\epsilon}(v)$ are true:
\begin{itemize}
  \item[1.] For each pair of straight-line edges $e_1$ and $e_2$ that do not share a common vertex, $T^{\delta}(e_1) \cap T^{\delta}(e_2)=\emptyset$.
  \item[2.] For each pair of distinct straight-line edges $e_1$ and $e_2$ sharing a common vertex $v$, $(T^{\delta}(e_1) \cap T^{\delta}(e_2)) \subset C^{\epsilon}(v)$, i.e., there is no intersection between $\delta$-tubes outside the $\epsilon$-cubes.
  \item[3.] For a vertex $v$ and a straight-line edge $e$ not containing $v$, $C^{\epsilon}(v)\cap T^{\delta}(e)=\emptyset$.
\end{itemize}
\end{obs}

Now, we bound the value of $\delta$. For bounding the value of $\delta$, we proceed in a similar way as in Lemma \ref{lem:deltavalue} and we get that $\delta=\frac{\mathcal{C'}}{r^2n^2}$, where $\mathcal{C'}$ is a constant and $r$ is the chromatic number of the graph. Next we refine the grid sufficiently until the length of each grid edge becomes $\frac{\mathcal{C'}}{100 \cdot r^2n^2}$. This is done to ensure that there are enough grid points within $T^{\delta}(e)$ to convert the edge $e$ into a Manhattan path inside $T^{\delta}(e)$. The conversion of a straight-line edge $e$ connecting $u$ and $v$ into a Manhattan path $M(e)$ between $u$ and $v$ is done in the similar way as in Lemma~\ref{lem:stline_to_Manhat} in section~\ref{sec:embedding}. Also the conversion of the straight-line edge is done in such a way that it avoids the edges of the cube $C^{\epsilon}(v)$. Hence, we have obtained an embedding of an $r$-chromatic graph on a refined grid of size $O(r^4n^3)\times O(r^3n^4)\times O(r^4n^4)$, where each edge $e$ is represented by a Manhattan path $M(e)$.


\begin{figure}[!h]

\centering
\includegraphics[scale=0.4]{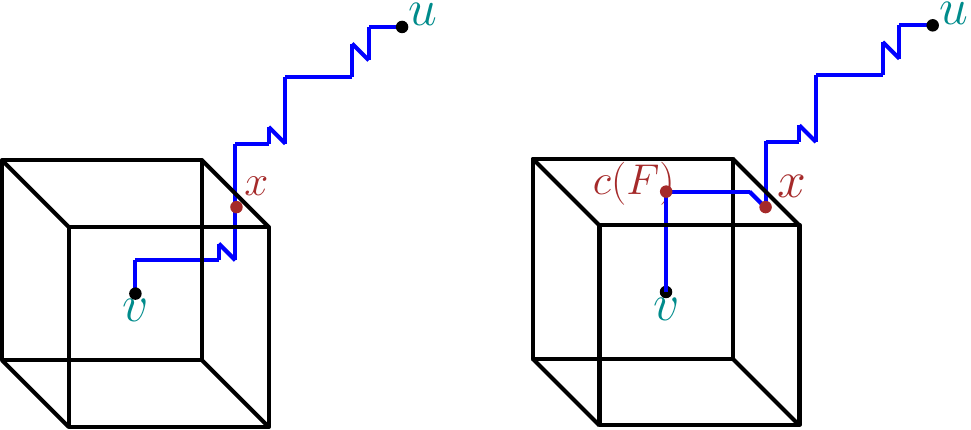} \\
\caption{Modification of a Manhattan path inside $C^{\epsilon}(v)$.\\}
\label{fig:3d}
\centering

\end{figure}
Next, we modify the Manhattan paths inside $C^{\epsilon}(v)$ in the way as shown in Figure~\ref{fig:3d}. For each square face $F$ of $C^{\epsilon}(v)$, let $c(F)$ be the point of intersection of the diagonals of $F$. Each Manhattan path $M(e)$ enters the cube $C^{\epsilon}(v)$ through a point, say $x$, on some square face $F$ of $C^{\epsilon}(v)$. Now, alter the portion of the Manhattan path $M(e)$ inside $C^{\epsilon}(v)$ as follows: (i) take any Manhattan path from $x$ to $c(F)$ and then (ii) take the straight-line path from $c(F)$ to $v$. We do this modification for every Manhattan path inside $C^{\epsilon}(v)$. Now we place obstacles on the edges of $C^{\epsilon}(v)$ and everywhere inside the cube $C^{\epsilon}(v)$ except square faces and the axis parallel straight-lines containing $v$. We do this modification for every cube $C^{\epsilon}(v)$. The Manhattan path $M(e)$ between $u$ and $v$ is now modified to another path containing the altered path inside $C^{\epsilon}(u)$, followed by the portion of $M(e)$ outside $C^{\epsilon}(u)$ and $C^{\epsilon}(v)$, and finally the altered path inside $C^{\epsilon}(v)$. Next we place obstacles on each empty grid point throughout the grid. Note that the modified paths are also Manhattan paths. Hence, we have the following theorem.

\begin{theo}\label{theo:3d}
Every $r$-colorable graph admits a $(\mathbb{Z}^{3}, \ell_1)$ obstacle representation on an $O(r^4n^3)\times O(r^3n^4)\times O(r^4n^4)$ grid.
\end{theo}

\subsection{Embedding in a horizontal strip}
\label{sec:horizontal_grid}
\noindent
In this section, we study the $\ell_1$-obstacle representation in $\Z^2$ of a graph $G$ inside a \emph{horizontal strip}. A \emph{horizontal strip} is a grid where the $y$-coordinates are bounded but the $x$-coordinates can be an arbitrary integer. Note that an $\ell_1$-obstacle representation in a horizontal strip for a graph implies its $\ell_1$-obstacle representation in $\Z^2$. But the converse is not true. For example, $C_n$ (cycle on $n$ vertices) and $K_n$ (complete graph on $n$ vertices), for $n>4$, admit $\ell_1$-obstacle representation in $\Z^2$ but not $\ell_1$-obstacle representations in horizontal strips containing only two rows.

We present a compression technique to show that if a graph $G$ admits an $\ell_1$-obstacle representation in a horizontal strip, then it admits an $\ell_1$-obstacle representation in a grid whose size is linear in the number of vertices of $G$ (assuming the height of the horizontal strip is constant). In effect, the importance of the result is in showing that if a graph admits an $\ell_1$-obstacle representation in a horizontal strip of height $b$, then it admits an $\ell_1$-obstacle representation in a polynomial-sized grid of size $b \times O(b^3n)$.  \remove{The case of \emph{vertical strip} can be argued in a similar way.}

\remove{Once we know that a graph admits an $\ell_1$-obstacle representation, the next obvious thing is to look for an $\ell_1$-obstacle representation which takes minimum possible grid size. The study of $\ell_1$-obstacle representation inside a \emph{horizontal strip} is a step towards this direction. We present a compression technique to show that if a graph admits an $\ell_1$-obstacle representation in a horizontal strip, then it admits an $\ell_1$-obstacle representation in a finite grid.}

\begin{theo}\label{lemma:finitehorizontalembed}
Let $G$ admit an $\ell_1$-obstacle representation in a horizontal strip of height $b$. Then $G$ has an $\ell_1$-obstacle representation on a finite grid of size $b \times O(b^3n)$.
\end{theo}

\begin{proof}
Let $embd(G)$ be the $\ell_1$-obstacle representation of $G$ in a horizontal strip of height $b$. For ease of exposition, we prove the theorem for the case where vertices of $G$ have different $x$-coordinates in $embd(G)$. The same argument holds for the case where some vertices have the same $x$-coordinates. We say that two vertices $v_1$ and $v_2$ are consecutive if there is no other vertex whose $x$-coordinate lies between the $x$-coordinates of $v_1$ and $v_2$. If for any two consecutive vertices in $embd(G)$, the difference between their $x$-coordinates is less than $O(b^3)$, then the theorem is immediate. Hence, we aim to prove, by a compression argument, that there exists an $\ell_1$-obstacle representation of $G$ in a horizontal strip of height $b$ such that
\begin{itemize}
\item the consecutive vertices of $G$ in $embd(G)$ remain consecutive in the new representation, and
\item the difference between $x$-coordinates of two consecutive vertices is $O(b^3)$.
\end{itemize}


For the rest of the proof, we focus on a portion $T$ of $embd(G)$ between two consecutive vertices of $G$. We modify each of these portions to get a different representation of $G$ in the same horizontal strip. The modification is as follows.
Let $l$ and $r$ be the starting and ending vertical lines of $T$, respectively. Note that both $l$ and $r$ have $b$ grid points each. In $embd(G)$, there may be multiple Manhattan paths from a grid point of $l$ to a grid point of $r$. Let ${\cP}(i,j)$ denote the set of all Manhattan paths from the $i$-th grid point of $l$ to the $j$-th grid point of $r$. In the new representation, we only maintain a single path for each ${\cP}(i,j)$ and color those paths. In $T$, we retain all the colored paths and put obstacles everywhere else. Let us denote this new representation by $embd'(G)$. Note that, in $embd'(G)$, the connectivity between $i$-th grid point of $l$ and $j$-th
grid point of $r$ is maintained, if ${\cP}(i,j)$ is non-empty. Also, putting obstacles does not create any unwanted connectivity. Hence, we have the following claim:

%


\begin{cl}
The new representation $embd'(G)$ is an $\ell_1$-obstacle representation of $G$.
\end{cl}

Next we calculate the total number of \emph{bend points} of $embd'(G)$.
A \emph{bend point} of $embd'(G)$ is a grid point on some colored path $\cP$
where $\cP$ changes direction. Since the height of the grid is bounded by $b$, the number of bend points on a colored path is at most $2b-2$. Moreover, every bend point is on some colored path of some $\cP(i,j)$. Since there are at most $O(b^2)$ non-empty $\cP(i,j)$s, the total number of bend points is $O(b^3)$ in $T$.

Two bend points $b_1$ and $b_2$ are consecutive if there is no other bend point whose $x$-coordinate lies between the $x$-coordinates of $b_1$ and $b_2$. The following claim shows that the horizontal distance between two consecutive bend points need not be arbitrarily large.
\begin{figure}[h]
\begin{center}
\includegraphics[width=8.0cm]{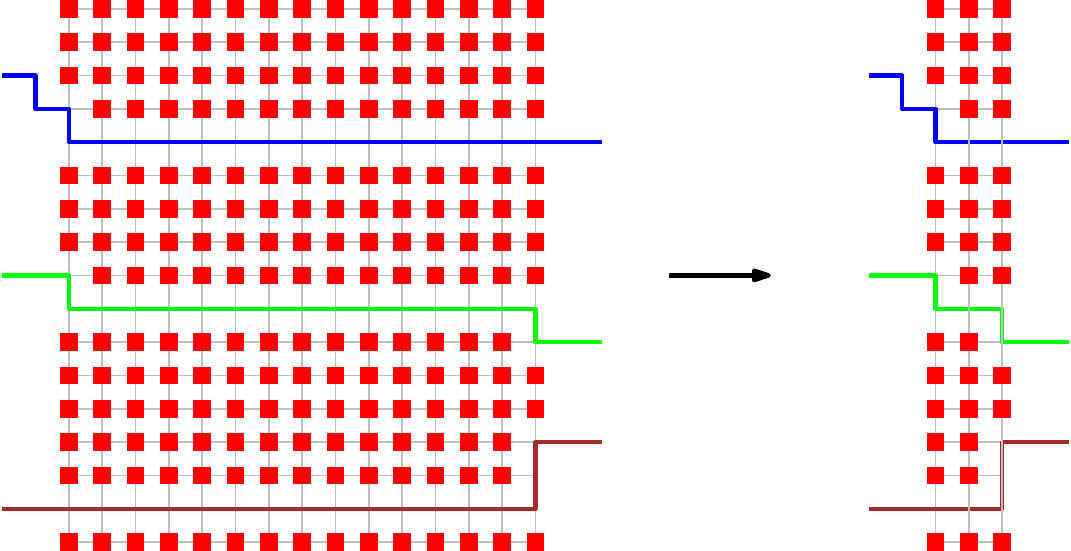}
\caption{Illustration of compression}
\label{fig:compression}
\end{center}
\end{figure}

\begin{cl}
The representation $embd'(G)$ can be modified to another representation $embd''(G)$ where the number of vertical lines between two consecutive bend points is constant.
\label{cl:embed2}
\end{cl}
\begin{proof}
Note that in $embd'(G)$ all the colored paths that cross the section between two consecutive bend points are horizontal.\remove{That is, all the paths start at some $y$-coordinate of the left vertical grid line of the section and end at the same $y$-coordinate of the right vertical grid line of the section.} Moreover, in $embd'(G)$, there are obstacles everywhere except colored paths. Therefore, the vertical grid lines between two consecutive bend points can be compressed to only three vertical grid lines such that the first and the third vertical grid line is identical with the first and last vertical grid line of the section. We keep any one of the intermediate vertical grid lines of $embd'(G)$ as the middle grid line. Repeating this process for every consecutive pair of bend points, we have an $embd''(G)$ where the number of vertical lines between two consecutive bend points is constant.
\end{proof}


So, in the representation $embd''(G)$, the total number of bend points is $O(b^3)$ and between two consecutive bend points, we have only one vertical grid line. Hence, there are $O(b^3)$ vertical grid lines between any two consecutive vertices. Therefore, $embd''(G)$ is of the size $b \times O(b^3n)$.
\end{proof}

%% file: obstacle-non-existence.tex
\section{Nonexistence of grid obstacle representation}
\label{sec:nonexistence}
\noindent
In this section, we show that, in $\Z^2$, not every graph admits a grid obstacle representation\footnote{The proof of this result was first given by J\'{a}nos Pach~\cite{JanosPach2016} who came to know of the grid obstacle representation from our manuscript. However, the proof presented here is different and is based on the suggestion of an anonymous reviewer of an earlier version of this paper. 
}. Let $G=(V,E)$ be a $C_4$-free\footnote{In this case $C_4$-free means no subgraph of $G$ is $C_4$.} graph on more than $20$ vertices having at least $8n-19$ edges. The existence of such a graph is known, see for example~\cite{Furedi1961}. We will show that $G$ has no grid obstacle representation in $\Z^2$.

A graph is called \emph{quasiplanar} if it admits a drawing in a plane such that there does not exist three pairwise crossing edges. The maximum number of edges in a quasiplanar graph is $8n-20$~\cite{AcKT07}. So, the graph $G$ considered above is not quasiplanar.

\begin{theo}\label{theo:nonexistential}
There exists a graph that does not admit a grid obstacle representation in $\Z^2$.
\end{theo}
\begin{proof}
Let $G=(V,E)$ be a non-quasiplanar, $C_4$-free graph on more than $20$ vertices having at least $8n-19$ edges. Assume that $G$ admits a grid obstacle representation with the mapping $f: V \rightarrow P$. Hence, for each $e= uv \in E$, there is an $\ell_1$-path, say ${\cP}_e$, from $f(u)$ to $f(v)$ such that it does not encounter any obstacle. Since $G$ is not quasiplanar, there exist three disjoint edges $e_1, e_2$ and $e_3$ in $E$ such that ${\cP}_{e_1}, {\cP}_{e_2}, {\cP}_{e_3}$ have a pairwise crossing.

Consider, for each $i\in \{1,2,3\}$, the path ${\cP}_{e_i}$ to be going from a point $u_i$ to another point $v_i$ such that the $x$-coordinates of $u_i$ is smaller than that of $v_i$. Except for the case where both the end points of an $\ell_1$-path have same $y$-coordinates, all the paths are either increasing (going from a point having smaller $y$-coordinate to a point having larger $y$-coordinate) or decreasing (going from a point having larger $y$-coordinate to a point having smaller $y$-coordinate). So, there must exist two paths among ${\cP}_{e_1}, {\cP}_{e_2}, {\cP}_{e_3}$ that are either both non-decreasing or both non-increasing. Without loss of generality, assume that ${\cP}_{e_1}$ and ${\cP}_{e_2}$ are non-decreasing. Let $e_1 = u_1v_1, e_2= u_2v_2$ and $p \in \Z^2$ be the grid point where ${\cP}_{e_1}$ and ${\cP}_{e_2}$ cross each other. Note that there exists an $\ell_1$-path between $f(u_1)$ and $f(v_2)$ through $p$. The path first follows $P_{e_1}$ up to the point $p$ and then it follows $P_{e_2}$. Similarly, there is an $\ell_1$ path between $f(u_2)$ and $f(v_1)$ through $p$. This implies that both the edges $u_1v_2, u_2v_1 \in E$. Note that we have a $C_4$ as a subgraph on the vertex set $\{u_1,u_2,v_1,v_2\}$, which is a contradiction.
\end{proof}

%% file: obstacle-nphard.tex
\section{Hardness results}
\label{app:hardness}
\noindent
Here, we study the following problem of $\ell_1$-obstacle representability of a graph. \remove{where the vertices are mapped to a specific set of a {\em polynomial grid}, i.e.,
a grid whose size is polynomial in the size of the graph under consideration.
The formal problem statement is as follows:}

\medskip

\noindent\underline{$\ell_1$-obstacle representability on a given point set ($\ell_1$-OEPS)}
\begin{description}
\item[\noindent\emph{Instance:}] A graph $G=(V,E)$ and a subset $S$ of a polynomial sized (polynomial in $\size{V}$) grid points with $|S|=|V|$ 

\item[\noindent\emph{Question:}] Does there exist an $\ell_1$-obstacle representation of $G$ such that the vertices of $G$ are mapped to $S$?
\end{description}

%

We show that $\ell_1$-OEPS is NP-complete for subdivision of simple non-Hamiltonian planar cubic graphs. The reduction is from a restricted version of \emph{geodesic point set embeddability (GPSE)} problem. The problem is whether a planar graph has a Manhattan-geodesic drawing, i.e., a drawing in which edges are Manhattan paths between the end points, such that the vertices are embedded onto a given set of points $S$. In the restricted version of GPSE problem, the given point set $S$ is partitioned into three specific sets, say $P_0, P_1$ and $P_2$, where $P_0 = \{(-j,0)|j=0,1,\ldots, 2n-2\}$, $P_1 = \{(j,nj)|j=1,2,\ldots, k_1\}$, and $P_2 = \{(j,-nj)|j=1,2,\ldots, k_2\}$ with $k_1+k_2=n/2+1$. This restricted version is known to be NP-complete \cite{Katz:2010} for subdivision of simple non-Hamiltonian planar cubic graphs, where each edge is subdivided exactly once. The formal problem statement is as follows:

\medskip

\noindent\underline{Restricted Manhattan-geodesic embeddability \remove{on a given point set} ($(P_0,P_1,P_2)$-GPSE)}
\begin{description}
\item[\noindent\emph{Instance:}] A planar graph $G=(V,E)$ and three specific sets $P_0,P_1$ and $P_2$ of grid points, as mentioned above, with $|P_0|+|P_1|+|P_2|=|V|$.

\item[\noindent\emph{Question:}] Does there exist a Manhattan-geodesic embedding of $G$ such that the vertices of $G$ are mapped to $P_0,P_1$ and $P_2$?
\end{description}

\begin{theo}
The problem $\ell_1$-OEPS is NP-complete for subdivision of simple non-Hamiltonian planar cubic graphs.
\end{theo}

\begin{proof}
Note that a certificate of $\ell_1$-OEPS is a mapping $f$ from $V$ to $S$ plus a set of obstacles $\mathcal{O}$. Since the grid is of polynomial size,
the number of obstacles is also polynomial. It is easy to see that given a certificate, we can check in polynomial time whether $G$ realizes an
$\ell_1$-obstacle representation by invoking shortest path algorithm $O(n^2)$ times. Hence, $\ell_1$-OEPS is in NP.



%


Let $G'=(V',E')$ be an instance of $(P_0,P_1,P_2)$-GPSE, i.e., $G'$ is a subdivision of a simple non-Hamiltonian planar cubic graph, say $G=(V,E)$, where each edge of $G$ is subdivided exactly once. Note that $|V|=n$ is even and $|E|=3n/2$. Therefore, $|V'|=5n/2$ and $|E'|=3n$. For some $k_1, k_2$ with $k_1+k_2=n/2+1$, let
$P_0 = \{(-j,0)|j=0,1,\ldots, 2n-2\}$,
$P_1 = \{(j,nj)|j=1,2,\ldots, k_1\}$,
$P_2 = \{(j,-nj)|j=1,2,\ldots, k_2\}$,
$P'_1 = \{(2j,2nj)|j=1,2,\ldots, k_1\}$, and
$P'_2 = \{(2j,-2nj)|j=1,2,\ldots, k_2\}$.
Let $S=P_0\cup P_1\cup P_2$ and $S'=P_0\cup P'_1\cup P'_2$. For such $k_1, k_2$ with $k_1+k_2=n/2+1$, the following claim shows that $\ell_1$-OEPS is NP-complete for subdivision of simple non-Hamiltonian planar cubic graphs.

\begin{figure}[h]
	\begin{center}
		\includegraphics[width=7cm]{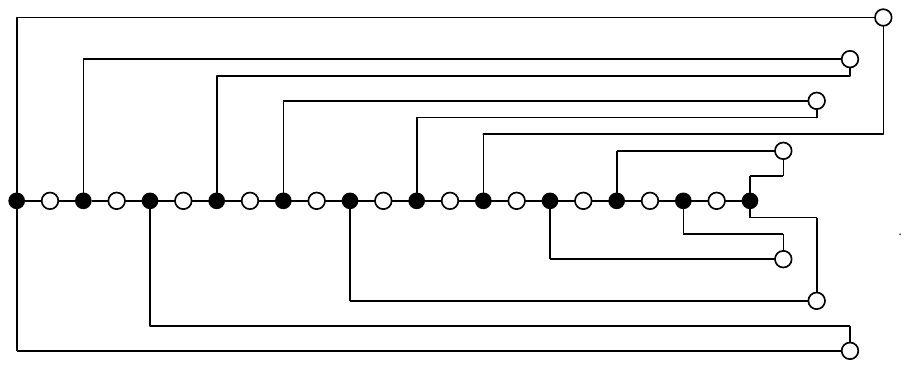}
  		\captionof{figure}{Geodesic embedding of $G'$.
  		The solid circles correspond to the vertices in $G$ and
   		the hollow circles correspond to the new vertices introduced by the subdivision.}
  		\label{fig:hardness1}
	\end{center}
\end{figure}

\begin{figure}[h]
	\begin{center}
		\includegraphics[width= 7cm]{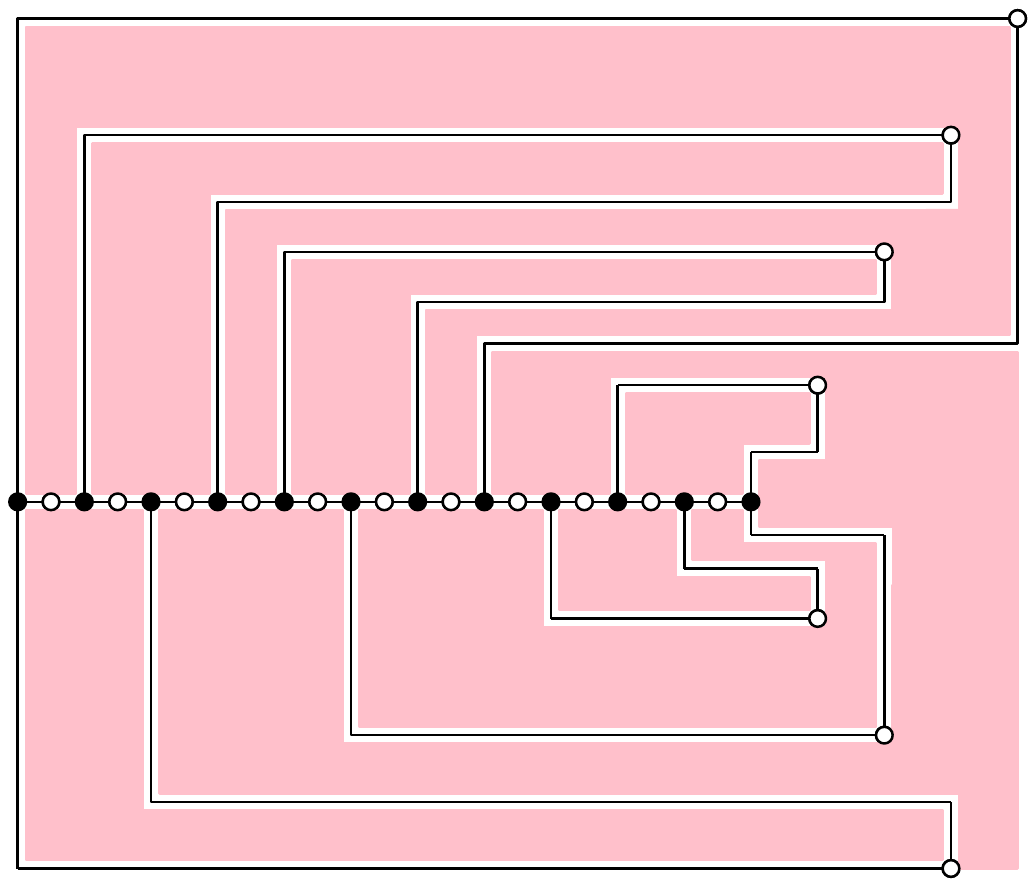}
  		\captionof{figure}{$\ell_1$-obstacle representation of $G'$.}
  		\label{fig:hardness2}
	\end{center}
\end{figure}

\begin{cl}\label{claim-for-NP-hardness}
$G'$ has a Manhattan-geodesic embedding on $S$ if and only if $G'$ has an $\ell_1$-obstacle representation on $S'$. 
\end{cl}

\begin{proof}
Let $G'$ have a Manhattan-geodesic embedding on $S$ (see Figure \ref{fig:hardness1}). From the proof of NP-hardness of GPSE \cite{Katz:2010}, it follows that only vertices with degree $2$ can be mapped to $P_1$ and $P_2$ and vertices of degree $2$ and degree $3$ alternate on $P_0$ from left to right. We can get an $\ell_1$-obstacle representation of $G'$ in the following way:

First, insert a grid row between every pair of consecutive rows in the grid and then insert a grid column between every pair of consecutive columns in the half where the $x$-coordinate is non-negative. After that, place obstacles everywhere in the grid except the paths given by the Manhattan-geodesic embedding of $G'$ (see Figure \ref{fig:hardness2}). Note that in this embedding, no two vertical segments of two distinct edges are next to each other in the $x<0$ half-plane because all such vertical segment uses $x=a$ line, where $a$ is even. Hence, this is an $\ell_1$-obstacle representation of $G'$ on $S'$.




Conversely, let $G'$ have an $\ell_1$-obstacle representation on $S'$. First we make some claims regarding the $\ell_1$-obstacle representation of $G'$ on $S'$, the proofs of which are done later.

\begin{cl}\label{Claim:deg2}
All the vertices that are mapped to $P'_1$ and $P'_2$, are of degree $2$.
\end{cl}

\begin{cl}\label{Claim:nonintersecting}
No two paths between disjoint pairs of vertices share a common grid point in the $\ell_1$-obstacle representation of $G'$ on $S'$.
\end{cl}

\begin{figure}[h]
	\begin{center}
		\includegraphics[height= 15cm, angle= -90]{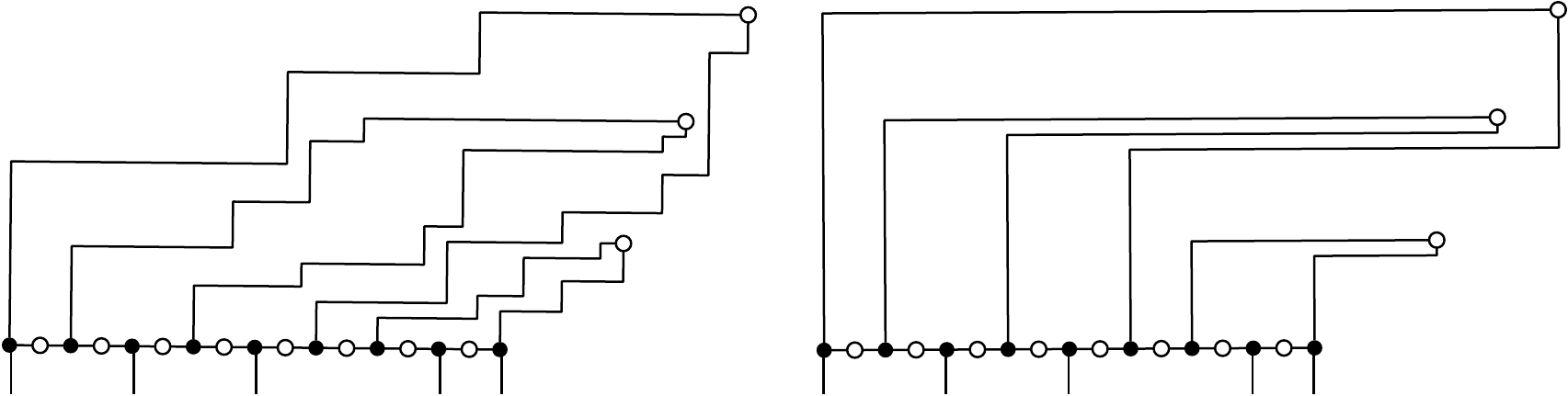}
  		\captionof{figure}{Modification of the Manhattan paths to obtain new $\ell_1$-obstacle representation.}
  		\label{fig:hardness222}
	\end{center}
\end{figure}

By Claim \ref{Claim:nonintersecting}, an $\ell_1$-obstacle representation on $S'$ is a Manhattan geodesic embedding on $S'$. Given such an $\ell_1$-obstacle representation of $G'$ on $S'$, we modify some of the Manhattan paths of the $\ell_1$-obstacle representation to get an $\ell_1$-obstacle representation of $G'$ on $S'$ such that all the paths follow grid lines of the form $x=a$ or $y=b$, where $a$ and $b$ are even and no two paths share any grid point. We explain this modification process only for the Manhattan paths between vertices of $P'_1$ and $P_0$ (see Figure \ref{fig:hardness222}). The Manhattan paths between the vertices of $P'_2$ and $P_0$ will be modified similarly. Note that the unique Manhattan paths between vertices of $P_0$ can not be modified. For each $j\in \{1,2,\ldots, k_1\}$, the vertex $(2j, 2nj)$ is of degree 2 by Claim \ref{Claim:deg2}. The two Manhattan paths incident on the vertex are modified as follows: let the degree $2$ vertex $v$ at $(2j,2nj)$ has Manhattan paths to two degree $3$ vertices $u_1$ and $u_2$ on $P_0$. Let the $x$-coordinates of $u_1$ and $u_2$ be $-2t_1$ and $-2t_2$, respectively, with $t_1>t_2$.
\begin{enumerate}
\item First we replace the Manhattan path between $v$ and $u_1$ by the path from $v$ to $(-2t_1, 2nj)$ along $y=2nj$ and then to $u_1$ along $x=-2t_1$. Note that it is a Manhattan path with only one bend.

\item The new path between $u_2$ and $v$ is described as follows: from $u_2$, we first take vertical path upwards along $x= -2t_2$ till the point $(-2t_2, s)$, where $s$ is even and there is a newly formed Manhattan path passing through $(-2t_2, s+2)$. Next we take the path from $(-2t_2, s)$ to $(2j,s)$ along $y=s$. And finally it reaches $(2j,2nj)$ from $(2j,s)$ along $x=2j$. Note that it is a Manhattan path with exactly two bends.
\end{enumerate}

Note that, in this new $\ell_1$-obstacle representation, every vertical segment of paths in $x<0$ half-plane are not next to each other because they pass through $x=a$ line, where $a$ is even. Once we have this new $\ell_1$-obstacle representation of $G'$ on $S'$, if we delete all the rows of the form $y=b'$, where $b'$ is odd, and delete all the columns of the form $x=a'$, where $a'$ is any positive odd number, then we have a Manhattan-geodesic embedding of $G'$ on $S$.
\end{proof}

Hence, the $\ell_1$-OEPS is NP-complete for subdivision of simple non-Hamiltonian planar cubic graphs.
\end{proof}

\noindent \emph{Proof of Claim \ref{Claim:deg2}:} Note that, out of the $n$ vertices of degree $3$, at least $n-1$ vertices of degree $3$ have to be mapped in $P_0$ because otherwise two vertices of degree $2$ would be consecutive in $P_0$, which is a contradiction to
the fact that there is no edge between degree $2$ vertices in $G'$. Hence, at most one degree $3$ vertex can be mapped to some point in $P'_1$ or $P'_2$. Further note
that, if there is a vertex of degree $3$, say $v$, that is mapped to some point
in $P'_1$ or $P'_2$, then the vertices mapped to $(0,0)$ and $(-2n+2,0)$ is of degree $2$ by the pigeonhole principle. In that case, these degree $2$ vertices must be adjacent to $v$ because degree $2$ vertices are adjacent to degree $3$ vertices in $G'$ and $v$ is the only degree $3$ vertex available (other than those at $(-1,0)$ and $(-2n+3,0)$).
This implies that $G'$ has a cycle of length $2n$ with all of the degree $3$ vertices,
which is a contradiction to the fact that $G$ is non-Hamiltonian.
Hence, all the vertices that are mapped to $P'_1$ and $P'_2$, are of degree $2$.

\noindent \emph{Proof of Claim \ref{Claim:nonintersecting}:} Let $p_1$ and $p_2$ be two unblocked Manhattan paths in the $\ell_1$-obstacle representation of $G'$ on $S'$ from $x_1$ to $y_1$ and from $x_2$ to $y_2$ respectively, where $x_1,x_2\in P_0$ with at least one of them, say $x_1$, does not belongs to  $\{ (0,0), (-2n + 2, 0) \}$, and $y_1,y_2\in P'_1$. If $p_1$ and $p_2$ share any grid point, then there would be a Manhattan path between $x_1,y_2$. This implies that the degree of $x_1$ is $4$, which is a contradiction. The argument also holds if $y_1,y_2\in P'_2$. Also, if $y_1\in P'_1$ and $y_2\in P'_2$, then the paths $p_1$ and $p_2$ cannot share any grid point. Now for the case when $x_1, x_2 \in \{ (0,0), (-2n + 2, 0) \}$, let $x_1'$ and $x_2'$ be the other end points of the Manhattan paths incident to $y_1$ and $y_2$, respectively. With similar arguments, it follows that the degree of $x'_1$ and $x'_2$ is $4$, which is a contradiction.

%% file: obstacle-conclusion.tex
\section{Conclusion}
\label{sec:conclusion}
\noindent
We have studied the grid obstacle representation of graphs. A generalized version of this, namely \emph{geodesic obstacle representation} has been very recently studied in \cite{GeodesicBose}. In this article, our main focus has been on the existential question of grid obstacle embedding of graphs in polynomial sized grids. We have proved that planar graphs admit grid obstacle representation in grids of size $O(n^4)\times O(n^4)$ in $\mathbb{Z}^2$. Motivated by our definition of grid obstacle representation, Biedl and Mehrabi showed recently that planar graphs admit grid obstacle representations in grids of size $O(n)\times O(n)$ in $\mathbb{Z}^2$ \cite{Biedl2018}. As planar graphs admit grid obstacle representation and there exist graphs that do not, a pertinent question is to characterize the graphs that admit grid obstacle representation in $\Z^2$. There are two associated optimality problems --- given a graph $G$ that admits an $\ell_1$-obstacle representation, find the $\ell_1$-obstacle number and the minimum grid size for an $\ell_1$-obstacle representation. We highlight some interesting problems in this area. 

\begin{prob}
Characterize graphs that admit grid obstacle representation in $\mathbb{Z}^2$.
\end{prob}

There are mainly two optimization problems associated with the $\ell_1$-obstacle representation of a graph. The problems are as follows:

\begin{prob}
Given a graph $G$ that admits an $\ell_1$-obstacle representation, find the $\ell_1$-obstacle number of $G$ on $\mathbb{Z}^2$.
\end{prob}

\begin{prob}
Given a graph $G$ that admits an $\ell_1$-obstacle representation, find the minimum grid size for $\ell_1$-obstacle representation of $G$ on $\mathbb{Z}^2$.
\end{prob}